\documentclass[prd,aps,floats,preprintnumbers,preprint,nofootinbib]{revtex4}
\usepackage[dvipdfm]{graphicx,color}

\usepackage{amssymb,amsmath,amsfonts,amsbsy}

\newcommand{\mpl}{m_{\rm Pl}}
\newcommand{\calA}{{\cal A}}
\newcommand{\calB}{{\cal B}}
\newcommand{\calC}{{\cal C}}
\newcommand{\calE}{{\cal E}}
\newcommand{\calF}{{\cal F}}
\newcommand{\calG}{{\cal G}}
\newcommand{\calH}{{\cal H}}
\newcommand{\calL}{{\cal L}}
\newcommand{\calR}{{\cal R}}

\begin{document}

\preprint{YITP-10-5}

\title{A complete analysis of
\\
linear cosmological perturbations
\\
in Ho\v{r}ava-Lifshitz gravity}

\author{
 Jinn-Ouk Gong,$^{a}$\footnote{jgong$\mbox{@}$lorentz.leidenuniv.nl}
 Seoktae Koh$^{b}$\footnote{steinkoh$\mbox{@}$sogang.ac.kr}
and
 Misao Sasaki$^{c}$\footnote{misao$\mbox{@}$yukawa.kyoto-u.ac.jp}
}

\affiliation{
 ${}^a$Instituut-Lorentz for Theoretical Physics,
Universiteit Leiden, 2333 CA Leiden, The Netherlands
 \\
 ${}^b$Center for Quantum Spacetime,
 Sogang University, Seoul 121-742, Republic of Korea
 \\
 ${}^c$Yukawa Institute for Theoretical Physics,
 Kyoto University, Kyoto 606-8502, Japan}

\begin{abstract}

We investigate the linear cosmological perturbations in Ho\v{r}ava-Lifshitz
gravity with a scalar field.
Starting from the most general expressions of the metric perturbations
as well as that of a canonical scalar field, we decompose the scalar, vector
and tensor parts of the perturbed action. By reducing the Hamiltonian,
we find that there are two independent degrees of freedom for the tensor
 perturbations while none for the vector perturbations.
For the scalar perturbations, the remaining number of degrees of freedom,
which are all gauge invariant, depends on whether the projectable
condition is applied or not: two when applied, with one of them being possibly a ghost, and one when not applied.
 For both cases, we lose the time reparametrization symmetry of any kind. 

\end{abstract}

\maketitle

\section{Introduction}

Recently, a candidate of an ultraviolet renormalizable theory of gravity
suggested by Ho\v{r}ava in Ref.~\cite{Horava:2009uw} has attracted a huge
 interest, which is inspired by an idea of Lifshitz in condensed matter
physics~\cite{lifshitz}. The essential point of this theory of gravity,
which is often called Ho\v{r}ava-Lifshitz (HL) gravity, is the
violation of the Lorentz invariance because of the anisotropic scaling
behaviours of the time and space coordinates with a dynamical critical
 exponent $z$,
\begin{align}
t \to & l^z t \, ,
\\
x^i \to & l x^i \, ,
\end{align}
where $z\geq1$. In the four dimensional space-time, HL gravity has
 an ultraviolet ``fixed'' point for $z=3$. This is possible in a special
 foliation of the constant time hypersurfaces. Then, the action,
which now contains higher order spatial derivatives of the metric,
is invariant under the foliation-preserving diffeomorphism,
\begin{align}
t \to & \tilde{t}(t) \, ,
\\
x^i \to & \tilde{x}^i(t,x^i) \, .
\end{align}
The functions of the former are called ``projectable''~\cite{Horava:2009uw}.

While it is still controversial whether HL gravity is a complete,
consistent theory~\cite{controversy,Blas:2009yd}, the cosmology of
HL gravity has been extensively studied~\cite{HLcosmology,Mukohyama:2009gg}. Several important properties have been clarified, e.g. in the ultraviolet limit, the scalar field perturbations
may produce a scale-invariant spectrum, insensitive to the expansion rate of
the universe~\cite{Mukohyama:2009gg}. However, before we address any properties
 of cosmological perturbations in HL gravity, we first of all
must clarify physical and gauge degrees of freedom. For this purpose,
it is desirable to formulate the cosmological perturbations in HL gravity
in a gauge-invariant manner as done in Einstein gravity~\cite{linear}.

Since the structure of the HL gravity is different from that of the conventional
Einstein gravity, this issue should be thoroughly addressed
and carefully analyzed.
Only after then we can solve the relevant equations of motion of the
relevant variables and study their observational significance.
In this paper, in the context of the linear
perturbation theory, we investigate the structure of the perturbed
action of HL gravity with a canonical scalar field,
and identify the gauge and physical degrees of freedom,
and spell out the perturbation equations. However, we do not
solve the equations.

The organization of this paper is as follows.
In Section~\ref{sec_perturbed_action},
we expand the action of HL gravity with a canonical scalar field
to quadratic order. Then we derive the background equations of motion from
the linear action, which are frequently used in the subsequent calculations.
Next, in Section~\ref{sec_ten_vec}, we focus on the tensor and
vector perturbations, and present their Hamiltonians.
We find that the results are structurally analogous to
Einstein gravity for tensor and vector perturbations.
In the next two sections, we study the scalar
perturbations without and with the projectable condition.
 In Section~\ref{sec_no_proj} we first consider the case without
the projectable condition.
We rewrite the action for the scalar perturbations
in the first order form, i.e. in the Hamiltonian form.
Then analyzing the constraint equations by computing the Poisson algebra,
we identify the gauge degree of freedom and find that there is
a single physical degree of freedom, just like in Einstein gravity.
However, the important difference is that there is no temporal gauge
degree of freedom in HL gravity.
In Section~\ref{sec_proj} we repeat the same procedure
with the projectable condition. We find that there are two
physical degrees of freedom, one from gravity and one from the scalar
field. Then we present the equations of motion for the two relevant variables.
Finally we conclude in Section~\ref{sec_conclusion}.
Some formulas used in Section~\ref{sec_no_proj} are summarized in the 
Appendices.

\section{Perturbed action}
\label{sec_perturbed_action}

\subsection{Gravity sector}

We first consider only the gravity sector of the HL theory.
 We begin with the Arnowitt-Deser-Misner metric~\cite{ADM}
\begin{equation}\label{ADM_metric}
ds^2 = -N^2d\eta^2 + \gamma_{ij}(N^id\eta + dx^i)(N^jd\eta + dx^j) \, ,
\end{equation}
where we include perturbations in the lapse function $N$, the shift vector
 $N_i$ and the induced spatial metric $\gamma_{ij}$ as
\begin{align}
N = & a(\eta)(1+A) \, ,
\\
N_i = & a^2(\eta)\calB_i \, ,
\\
\gamma_{ij} = & a^2(\eta)
\left[ \delta_{ij}
+ 2h_L\delta_{ij}
 + 2\left(\calE_{ij}-\frac{1}{3}\delta_{ij}\calE^k{}_k \right)\right] \, ,
\end{align}
respectively. In what follows we set $\calR\equiv h_L-\calE^k{}_k/3$.
Now, we try to write the HL gravity sector with perturbations up
 to second order. The action is written as
\begin{align}\label{HL_action}
S_\mathrm{HL} = & \int d^4x N\sqrt{\gamma} \left[ \frac{1}{\kappa^2}
\left(K^i{}_jK^j{}_i - \lambda K^2 \right) + \mu R\right.
\cr
&\hspace{30mm}
 \left.+ \alpha_1 R^{ij}R_{ij}
 + \alpha_2 R^2 +
\alpha_3 \frac{\epsilon^{ijk}}{\sqrt{\gamma}}R_{il}\nabla_jR^l{}_k
 + \alpha_4C^{ij}C_{ij}
+ \sigma \right] \, ,
\end{align}
where $\gamma$ is the determinant of $\gamma_{ij}$, $1/\kappa^2$ is the
coupling of kinetic sector of HL gravity, $K_{ij}$ is the extrinsic curvature
\begin{equation}
K_{ij} \equiv \frac{1}{2N}
 \left( \nabla_iN_j + \nabla_jN_i - \gamma_{ij}' \right) \, ,
\end{equation}
with $\nabla_i$ being a three dimensional covariant derivative,
$R$ and $R_{ij}$ are the Ricci scalar and the Ricci tensor
constructed from $\gamma_{ij}$, $C_{ij}$ is the Cotton tensor given by
\begin{equation}
C^{ij} = \frac{\epsilon^{ikl}}{\sqrt{\gamma}}\nabla_k \left( R^j{}_l -
\frac{1}{4}\delta^j{}_lR \right) \, ,
\end{equation}
and $\sigma$ is the remaining miscellaneous terms like a cosmological constant.
We can return to Einstein gravity by setting $\lambda=1$, $\kappa^2=2/\mpl^2$, $\mu=1/\kappa^2$ and the other parameters zero.

\subsubsection{Linear action}

After substituting the metric (\ref{ADM_metric}), into the
action (\ref{HL_action}) and expanding, the linear order pieces are
collected to give, with $\calH \equiv a'/a$ and
$\Delta\equiv\delta^{ij}\partial_i\partial_j$ being the spatial Laplacian,
\begin{align}\label{linear_HL_action}
\delta_1S_\mathrm{HL} = &
\int d^4x a^2 \left\{ \frac{1-3\lambda}{\kappa^2} \left[
6\calH\calR' - 3\calH^2(A-3\calR)
- 2\calH \left( \calB^i{}_{,i} - \calE^i{}_i' \right) +
3\calH^2\calE^i{}_i \right] \right.
\nonumber\\
& \left. \hspace{1.8cm} + \mu \left[ -4\Delta\calR + 2 \left( \calE^{ij}{}_{,ij} -
\Delta\calE^i{}_i \right) \right] + a^2\sigma \left( A+3\calR+\calE^i{}_i \right)
\right\} \, .
\end{align}
At this point, we can in general decompose $\calB_i$ and $\calE_{ij}$ into
 pure scalar, vector and tensor components as
\begin{align}
\label{decomposition_0i}
\calB_i = & B_{,i} + S_i \, ,
\\
\label{decomposition_ij}
\calE_{ij} = & E_{,ij} + F_{(i,j)} + h_{ij} \, ,
\end{align}
where $S_i$ and $F_i$ are transverse vectors,
and $h_{ij}$ is a transverse-traceless tensor,
\begin{equation}
S^i{}_{,i} = F^i{}_{,i} = h^i{}_i = h^i{}_{j,i} = 0 \, .
\end{equation}
Then we have at the moment total four, four and two independent
degrees of freedom for scalar, vector and tensor metric perturbations, respectively.
Later when we consider the perturbation of a canonical scalar field, we
have an additional degree of freedom for scalar perturbation so that its total
number becomes five.
Then, we are left with a relatively simple linear order action
\begin{equation}\label{HL_linear_action}
\delta_1S_\mathrm{HL}
= \int d^4x a^2 \left\{ \frac{1-3\lambda}{\kappa^2}
\left[ -3\calH^2A - 3\left(2\calH'+\calH^2\right)\calR \right]
 + a^2\sigma (A+3\calR) \right\} \, .
\end{equation}
Note that there are only scalar contributions to the linear action of
 the gravity sector.

\subsubsection{Quadratic action}

Collecting the second order terms, and decomposing the perturbations into
scalar, vector and tensor components, a lot of tensor and especially vector
 contributions disappear. Further, it can be found that there is no coupled
terms between scalar, vector and tensor metric perturbations. Hence,
as long as there is no mixing between different modes in the energy-momentum
tensor, we can separately consider each mode, like the decomposition
theorem in cosmological perturbation theory in Einstein gravity.
If we explicitly collect scalar, vector and tensor contributions,
we obtain the respective actions separately as
\begin{eqnarray}
\label{HL_2nd_action_scalar}
&&\delta_2S_\mathrm{HL}^{(s)}
\nonumber\\
&& = \int d^4x a^2 \left\{ \frac{1-3\lambda}{\kappa^2} \left( 3{\calR'}^2
+6\calH(\calR-A)\calR' + \frac{9}{2}\calH^2(A-\calR)^2
+\calH \left[2\calR' + 3\calH(\calR-A) \right]\Delta E \right.\right.
\nonumber\\
&& \hspace{0.5cm} - 2\calH\calR^{,i}B_{,i}
- 2\left[ \calR' + \calH(\calR-A) \right] \Delta \left( B- E' \right) -
\frac{3}{2}\calH^2B^{,i}B_{,i}  - 2\calH\Delta E
\Delta\left( B - E' \right)
\nonumber\\
&& \left. \hspace{0.5cm} + 2\calH B^{,i}\Delta E_{,i}
+ 4\calH E^{,ij}(B-E')_{,ij} + \frac{3}{2}\calH^2(\Delta E)^2 -
3\calH^2E^{,ij}E_{,ij} \right)
\nonumber\\
&& \hspace{0.5cm}+ \frac{1-\lambda}{\kappa^2} \left[
\Delta\left(B-E'\right) \right]^2
 - 2\mu (\calR+2A)\Delta\calR
 + \frac{6\alpha_1}{a^2}(\Delta\calR)^2 +
\frac{16\alpha_2}{a^2} (\Delta\calR)^2
\nonumber\\
&& \left. \hspace{0.5cm}
+ a^2\sigma \left[ - \frac{A^2}{2} + 3A\calR + \frac{3}{2}\calR^2 +
(A+\calR)\Delta E + \frac{B^{,i}B_{,i}}{2}
 + \frac{1}{2}(\Delta E)^2 - E^{,ij}E_{,ij}
\right] \right\} \, ,
\end{eqnarray}
\begin{eqnarray}
\label{HL_2nd_action_vector}
&&\delta_2S_\mathrm{HL}^{(v)}
\nonumber\\
&& = \int d^4x a^2 \left[ \frac{1}{\kappa^2} \left\{ (1-3\lambda) \left[
-\frac{3}{2}\calH^2S^iS_i + 2\calH S^i\Delta F_i + 2\calH F^{i,j}
\left( S_i - F_i'\right)_{,j}
- \frac{3}{2}\calH^2F^{i,j}F_{i,j} \right] \right.\right.
\nonumber\\
&& \left.\left. \hspace{2.8cm} + \frac{1}{2} \left( S^i-{F^i}' \right)^{,j}
\left(S_i-F_i' \right)_{,j} \right\}
 + \frac{1}{2}a^2\sigma \left( S^iS_i - F^{i,j}F_{i,j}
\right) \right] \, ,
\\
\label{HL_2nd_action_tensor}
&& \delta_2S_\mathrm{HL}^{(t)}
\nonumber\\
&& =\int d^4x a^2 \left\{ \frac{1}{\kappa^2}
 \left[ {h^{ij}}'h_{ij}' - (1-3\lambda)
\left( 4\calH h^{ij}h_{ij}' + 3\calH^2h^{ij}h_{ij} \right) \right] + \mu
h^{ij}\Delta h_{ij} \right.
\nonumber\\
&& \left. \hspace{2.2cm} + \frac{\alpha_1}{a^2}\Delta h^{ij}\Delta h_{ij}
+\frac{\alpha_3}{a^3} \epsilon^{ijk} \Delta h_{il} \Delta h^l{}_{k,j} -
\frac{\alpha_4}{a^4} \Delta h^{ij}\Delta^2h_{ij}
 - a^2\sigma h^{ij}h_{ij} \right\} \, .
\end{eqnarray}

\subsection{Matter sector}

We consider the matter action of a scalar field as
\begin{equation}
S_\mathrm{M} = \int d^4x N\sqrt{\gamma} \left[ \frac{1}{2N^2}
 \left( \phi' - N^i\phi_{,i}\right)^2 - Z(\phi) - V(\phi) \right] \, ,
\end{equation}
where
\begin{equation}
Z(\phi) = \sum_{n=1}^3 \xi_n \partial_i^{(n)}\phi\partial^{i(n)}\phi \, ,
\end{equation}
with $(n)$ denoting $n$-th spatial derivative.
In the $z=3$ HL gravity, $n$ is at most 3 as shown above. By setting $\xi_1=1/2$ and $\xi_2=\xi_3=0$, we can recover the matter action in Einstein gravity.
Expanding $\phi$ into background and perturbation as
\begin{equation}
\phi(\eta,\mathbf{x}) = \phi_0(\eta) + \delta\phi(\eta,\mathbf{x}) \, ,
\end{equation}
we can easily find that
\begin{align}\label{p_expansion}
& \frac{1}{2N^2} \left( \phi' - N^i\phi_{,i} \right)^2 - Z(\phi) - V(\phi)
\nonumber\\
& = \left( \frac{{\phi_0'}^2}{2a^2} - V_0 \right)
 + \left[ \frac{1}{a^2} \left( \phi_0'\delta\phi' - {\phi_0'}^2A \right)
- V_\phi\delta\phi \right]
\nonumber\\
& \hspace{0.5cm} + \frac{1}{2a^2} \left( {\delta\phi'}^2
- 4\phi_0'A\delta\phi' - 2\phi_0'\calB^i\delta\phi_{,i}
+ 4{\phi_0'}^2A^2 - {\phi_0'}^2\calB^i\calB_i \right) - \delta{Z}
 - \frac{1}{2}V_{\phi\phi}\delta\phi^2 \, ,
\end{align}
where
\begin{equation}
\delta{Z}
= \sum_{n=1}^3 \xi_n \partial_i^{(n)}\delta\phi\partial^{i(n)}\delta\phi \, .
\end{equation}
Below we will denote by $p_{(0)}$ the two background
terms in the first parentheses on the right hand side of (\ref{p_expansion}).

\subsubsection{Linear action}

First let us consider the linear action of the matter sector.
Combined with the gravity sector linear action (\ref{HL_linear_action}),
we can derive the background equations of motion which can be used to
further reduce the second order action.

The linear action is written as
\begin{align}\label{matter_linear_action}
\delta_1S_\mathrm{M} = & \int d^4x a^2 \left[ \phi_0'\delta\phi'
 - {\phi_0'}^2A - a^2V_\phi\delta\phi
+a^2p_{(0)} \left( A + 3\calR \right) \right] \, .
\end{align}
Now we can write the equations derived from the total linear action,
i.e. the sum of (\ref{HL_linear_action}) and (\ref{matter_linear_action}).
They are easily found as
\begin{align}
\label{friedmann_eq}
& \calH^2 =
-\frac{\kappa^2}{3(1-3\lambda)} \left[ {\phi_0'}^2
 - a^2 \left( \sigma+p_{(0)} \right) \right] \, ,
\\
\label{H_evolution}
& 2\calH' + \calH^2 =
\frac{\kappa^2}{1-3\lambda}a^2 \left( \sigma+p_{(0)} \right) \, ,
\\
\label{phi_evolution}
& \phi_0'' + 2\calH\phi_0' + a^2V_\phi = 0 \, .
\end{align}
 If we return to Einstein gravity by setting the parameters appropriately,
 (\ref{friedmann_eq}), (\ref{H_evolution}) and (\ref{phi_evolution})
give the Friedmann equation, the evolution equation of $\calH$, and the
equation of motion of $\phi_0$, respectively.
Note that we can combine (\ref{friedmann_eq}) and (\ref{H_evolution}) to
obtain another useful relation
\begin{equation}\label{useful}
\calH'-\calH^2 = \frac{\kappa^2}{2(1-3\lambda)}{\phi_0'}^2 \, .
\end{equation}

\subsubsection{Quadratic action}

We can straightforwardly write the second order matter action by collecting
quadratic terms of the matter action. Again scalar, vector and tensor
contributions are decoupled to give
\begin{align}
\label{matter_2nd_action_scalar}
\delta_2S_\mathrm{M}^{(s)}
= & \int d^4x a^2 \left\{ \frac{1}{2}{\delta\phi'}^2 -
\delta{Z} - \frac{1}{2}a^2V_{\phi\phi}\delta\phi^2 - 2\phi_0'A\delta\phi' -
\phi_0'B^{,i}\delta\phi_{,i} \right.
\nonumber\\
& + 2{\phi_0'}^2A^2 - \frac{1}{2}{\phi_0'}^2B^{,i}B_{,i}
 + \left( A + 3\calR + \Delta E\right)
\left( \phi_0'\delta\phi' - {\phi_0'}^2A - a^2V_\phi\delta\phi \right)
\nonumber\\
& \left. + a^2p_{(0)} \left[ \frac{3}{2}\calR^2 + 3A\calR
- \frac{1}{2}A^2 +(A+\calR)\Delta E
 + \frac{1}{2}B^{,i}B_{,i} + \frac{1}{2}(\Delta E)^2 - E^{,ij}E_{,ij}
\right] \right\} \, ,
\\
\label{matter_2nd_action_vector}
\delta_2S_\mathrm{M}^{(v)} = & \int d^4x a^2 \left[
-\frac{1}{2}a^2F^{i,j}F_{i,j}p_{(0)} + \frac{1}{2} \left( a^2p_{(0)} - {\phi_0'}^2
\right)S^iS_i \right] \, ,
\\
\label{matter_2nd_action_tensor}
\delta_2S_\mathrm{M}^{(t)} = & \int d^4x \left( -a^4 h^{ij}h_{ij}p_{(0)} \right) \, .
\end{align}

\subsection{Total quadratic action}

Having found the quadratic actions in the gravity and matter sectors,
we can now write
the full second order action of the system.

\subsubsection{Tensor quadratic action}

We first start with the tensor action since this is the simplest.
Summing the gravity sector~(\ref{HL_2nd_action_tensor}) and the
 matter sector~(\ref{matter_2nd_action_tensor}),
integrating by parts and using (\ref{H_evolution}),
the tensor quadratic action is reduced to
\begin{eqnarray}
\label{reduced_2nd_tensor_action}
\delta_2S^{(t)} &=& \int d^4x a^2
 \Biggl( \frac{1}{\kappa^2}{h^{ij}}'h_{ij}' + \mu h^{ij}\Delta h_{ij}
\cr
&&
\qquad
+ \frac{\alpha_1}{a^2}\Delta h^{ij}\Delta h_{ij}
 + \frac{\alpha_3}{a^3} \epsilon^{ijk} \Delta h_{il} \Delta h^l{}_{k,j}
 - \frac{\alpha_4}{a^4} \Delta h^{ij}\Delta^2h_{ij}
\Biggr) \, .
\end{eqnarray}
We need not manipulate this quadratic action any further to make it
simpler: tensor is by itself gauge invariant, and there is no gauge ambiguity.
Note that (\ref{reduced_2nd_tensor_action}) reduces to the
well known tensor quadratic action in Einstein gravity by appropriately setting the parameters,
\begin{equation}
\delta_2S_\mathrm{Einstein}^{(t)} = \int d^4x a^2 \frac{\mpl^2}{2} \left( {h^{ij}}'h_{ij}' + h^{ij}\Delta h_{ij} \right) \, .
\end{equation}

\subsubsection{Vector quadratic action}

Next, we consider the vector perturbations. From (\ref{HL_2nd_action_vector})
 and (\ref{matter_2nd_action_vector}), using the background
equations (\ref{H_evolution}) and (\ref{useful}), and integrating by parts,
the quadratic order action of the vector perturbations is written as
\begin{equation}\label{reduced_2nd_vector_action}
\delta_2S^{(v)} = \frac{1}{2\kappa^2} \int d^4x a^2
 \left( S^i-{F^i}' \right)^{,j} \left( S_i-F_i' \right)_{,j} \, .
\end{equation}
Note that unlike tensor or (as we shall see below) scalar, the quadratic
 vector action is the same as that in Einstein gravity.
Thus we expect that there will be no dynamical evolution of
the vector perturbations, and indeed that is the case.

\subsubsection{Scalar quadratic action}

Now we turn to the scalar quadratic action. After a number of manipulations
using the background equations, total derivatives and integrations by parts,
 we find the quadratic scalar action as
\begin{align}\label{reduced_2nd_scalar_action}
\delta_2S^{(s)}
= & \int d^4x a^2 \left\{ \frac{1-3\lambda}{\kappa^2}
 \left[ 3{\calR'}^2 - 6\calH A\calR' +
\left( \calH'+2\calH^2 \right) A^2
- 2\left( \calR' - \calH A \right) \Delta(B-E')
\right] \right.
\nonumber\\
& + \frac{1-\lambda}{\kappa^2} \left[ \Delta\left(B-E'\right) \right]^2
- 2\mu(\calR+2A)\Delta\calR
+ \frac{2}{a^2} \left( 3\alpha_1 + 8\alpha_2 \right)(\Delta\calR)^2
\nonumber\\
& \left. + \frac{1}{2}{\delta\phi'}^2 - \delta{Z} -
\frac{1}{2}a^2V_{\phi\phi}\delta\phi^2 - \phi_0'A\delta\phi'
- 3\phi_0'\calR'\delta\phi
 -a^2V_\phi A\delta\phi + \phi_0'\delta\phi\Delta(B-E') \right\} \, .
\end{align}
From this, we can write by setting the parameters appropriately,
\begin{align}
\delta_2S^{(s)}_\mathrm{Einstein}
= & \int d^4x a^2 \left\{ \frac{\mpl^2}{2} \left[ -6{\calR'}^2
+12\calH A\calR' - 2 \left( \calH' + 2\calH^2 \right)A^2
 - 2(\calR+2A)\Delta\calR \right]
\right.
\nonumber\\
& \hspace{1.8cm}
 + \frac{1}{2} \left( {\delta\phi'}^2 - a^2V_{\phi\phi}\delta\phi^2
 -\delta\phi^{,i}\delta\phi_{,i} \right)
+ \left[ \phi_0' \left( A'-3\calR' \right) -2a^2V_\phi A\right]\delta\phi
\nonumber\\
& \left. \hspace{1.8cm} + \left[ \phi_0'\delta\phi
 + 2\mpl^2 \left( \calR' - \calH A\right) \right] \Delta(B-E') \right\} \, ,
\end{align}
which is in agreement with the scalar quadratic action in
Einstein gravity~\cite{linear, reduction}.

\section{Hamiltonian reduction of tensor and vector Lagrangians}
\label{sec_ten_vec}

Now we are ready to reduce the phase space of HL gravity with a canonical
scalar field. First we consider the tensor and vector perturbations
which are much simpler than the scalar perturbations, which thus will
be separately discussed.

\subsection{Tensor perturbation}

Again, let us start with the simplest case of the tensor perturbations.
 From (\ref{reduced_2nd_tensor_action}), we can see that $h_{ij}$ is
the canonical variable and its conjugate momentum is
\begin{equation}\label{momentum_hij}
\Pi^{ij} \equiv \frac{\delta}{\delta{h}_{ij}'}\delta_2S^{(t)} = a^2
\frac{2}{\kappa^2}{h^{ij}}' \, .
\end{equation}
Then, (\ref{reduced_2nd_tensor_action}) can be now written
in the first order form as
\begin{align}
\delta_2S^{(t)} = &\int d^4x
\Bigl[ \Pi^{ij}h_{ij}' - \calH^{(t)} \Bigr]\,,
\\
\calH^{(t)}=&
\frac{\kappa^2}{4a^2}\Pi^{ij}\Pi_{ij} - a^2\mu h^{ij}\Delta h_{ij}
- \alpha_1\Delta h^{ij}\Delta h_{ij} -
\frac{\alpha_3}{a}\epsilon^{ijk}\Delta h_{il}\Delta h^l{}_{k,j}  +
\frac{\alpha_4}{a^2}\Delta h^{ij}\Delta^2h_{ij}\,.
\label{2nd_tensor_action:Hform}
\end{align}
As can be read, it is already in the form without any constraint.
 Thus the two independent degrees of freedom, with which we start,
are all physical and they can be interpreted as two polarizations of
the gravitational waves, as in Einstein gravity, except for the fact
that they no longer respect the local Lorentz invariance.
The solution of the equation of tensor perturbations we can derive
from (\ref{2nd_tensor_action:Hform}) can be found in Ref.~\cite{Takahashi:2009wc}.

\subsection{Vector perturbation}

Next we consider the vector quadratic action, (\ref{reduced_2nd_vector_action}).
Since only $F_i$ has a term quadratic in the time derivative,
the associated conjugate momentum exists only for $F_i$, which is given by
\begin{equation}
\Pi^i = \frac{\delta}{\delta F_i'}\delta_2S^{(v)}
 = \frac{a^2}{\kappa^2}\Delta \left( S^i-{F^i}' \right) \, .
\end{equation}
Then, the vector quadratic action is written as
\begin{align}
\label{vec_action_1st_order}
\delta_2S^{(v)} = & \int d^4x \left( \Pi^iF_i'
-\calH^{(v)} - S_i\Pi^i\right)\,,
\\
\calH^{(v)}=&-\frac{\kappa^2}{2a^2}\Pi_i\Delta^{-1}\Pi^i\, ,
\end{align}
where $\Delta^{-1}$ is the inverse Laplacian operator.
Now it is clear that $S_i$ is not a dynamical variable but plays
the role of a Lagrange multiplier. The equations of motion of $S_i$
impose the constraints,
\begin{equation}
\Pi^i = 0 \, .
\label{vectorconstraints}
\end{equation}
As clear from (\ref{vec_action_1st_order}), they commute with $\calH^{(v)}$.
Hence they are first class constraints, representing two vector type
gauge degrees of freedom.

Plugging the constraints (\ref{vectorconstraints}) back into the quadratic
action (\ref{vec_action_1st_order}) gives a vanishing Lagrangian.
Thus vector perturbations are found to be non-dynamical
as in the case of Einstein gravity.

\section{Scalar perturbation: Without projectable condition}
\label{sec_no_proj}

Given the scalar quadratic action (\ref{reduced_2nd_scalar_action}),
we can in principle proceed straightforwardly. However, as we have seen
 earlier, the projectable condition is necessary to keep the consistent
anisotropic scaling. For our case, it is applied to the
the 00-component of the metric perturbation $A$. That is,
the projectable condition implies $A$ is a function of only time,
$A = A(\eta)$, not a space-time dependent field. In this section,
let us first consider the case without the projectable condition.
 In this case the structure of the scalar action looks {\em superficially}
 similar to that in Einstein gravity, and it has been also
studied~\cite{noproj_pert}.
Many important properties are, however, found to be
very different from Einstein gravity as we shall see below.

\subsection{First order form of the Lagrangian}

The conjugate momenta from (\ref{reduced_2nd_scalar_action}) are
\begin{align}
\label{Pi_psi1}
\Pi^\calR = & a^2 \left\{ \frac{1-3\lambda}{\kappa^2} \left[ 6\left(
\calR'-\calH A \right) + 2\Delta(B-E') \right] + 3\phi_0'\delta\phi \right\} \, ,
\\
\label{Pi_deltaphi1}
\Pi^{\delta\phi} = & a^2 \left( \delta\phi' - \phi_0'A \right) \, ,
\\
\label{Pi_E1}
\Pi^E = & a^2 \Delta \left[ 2\frac{1-3\lambda}{\kappa^2}\left( \calR'-\calH A
\right) - 2\frac{1-\lambda}{\kappa^2}\Delta(B-E') - \phi_0'\delta\phi \right] \, .
\end{align}
Then,
 we can use (\ref{Pi_psi1}), (\ref{Pi_deltaphi1}) and (\ref{Pi_E1})
to write the derivatives of the canonical variables in terms of the
conjugate momenta. After some arrangement, we find
\begin{align}
\label{1st_order_L}
\calL_2^{(s)} = & \Pi^\calR\calR' + \Pi^{\delta\phi}\delta\phi'
 + \Pi^EE' - \calH^{(s)} -A\calC_A - B\calC_B\, ,
\\
\calH^{(s)} = & \frac{\kappa^2}{4a^2}
 \left[-\Pi^\calR\Delta^{-1}\Pi^E + \frac{3}{2}\left(\Delta^{-1}\Pi^E\right)^2 +
\frac{2}{\kappa^2}{\Pi^{\delta\phi}}^2
 + \frac{1-\lambda}{2(1-3\lambda)}{\Pi^\calR}^2\right]
 + \frac{\kappa^2}{2(1-3\lambda)}\phi_0'\Pi^\calR\delta\phi
\nonumber\\
& + \frac{3\kappa^2a^2}{4(1-3\lambda)}{\phi_0'}^2\delta\phi^2
 + 2a^2\mu\calR\Delta\calR
 -2(3\alpha_1+8\alpha_2)(\Delta\calR)^2 + \frac{a^2}{2} \left( 2\delta{Z} +
a^2V_{\phi\phi}\delta\phi^2 \right) \, ,
\\
\calC_A = & \calH\Pi^\calR + \phi_0'\Pi^{\delta\phi} + 4a^2\mu\Delta\calR
+ a^2 \left(3\calH\phi_0' + a^2V_\phi \right)\delta\phi \, ,
\\
\calC_B = & \Pi^E \, ,
\end{align}
where we have used the background equation (\ref{useful})
to eliminate terms proportional to $A^2$.

\subsection{Poisson algebra}

As can be read from the first order form Lagrangian (\ref{1st_order_L}),
 $A$ and $B$ appear linearly without any time derivative.
 Hence their coefficients constitute constraint equations,
$\calC_A = \calC_B = 0$.
We can easily find their Poisson brackets vanish,
\begin{equation}
\left\{ \calC_A,\calC_B \right\} = 0 \, ,
\end{equation}
and trivially
$\left\{ \calC_A,\calC_A \right\} = \left\{ \calC_B,\calC_B \right\} = 0$.
In the case of Einstein gravity, both $\calC_A=0$ and $\calC_B=0$ are
first class constraints. As we shall see shortly, however,
this is not the case in HL gravity.

Now let us consider the Poisson brackets  of the Hamiltonian $\calH^{(s)}$
with $\calC_A$ and $\calC_B$ to check the consistency of the constraint
equations $\calC_A = \calC_B = 0$ with the equations of motion.
First we can easily find
\begin{equation}
\left\{ \calH^{(s)}, \calC_B \right\} = 0 \, .
\end{equation}
For the Poisson bracket with $\calC_A$, after some calculations and introducing
\begin{equation}\label{deltaZ_deriv}
\delta{Z}_{\delta\phi} = \left(-\xi_1\delta\phi\Delta\delta\phi
 + \xi_2\delta\phi\Delta^2\delta\phi
 - \xi_3\delta\phi\Delta^3\delta\phi \right)_{,\delta\phi}
= -2\xi_1\Delta\delta\phi + 2\xi_2\Delta^2\delta\phi
 - 2\xi_3\Delta^3\delta\phi \, ,
\end{equation}
we find that
\begin{align}
\left\{ \calH^{(s)},\calC_A \right\}
= & \calC_A' - \calH\calC_A + \mu\kappa^2\calC_B+\calC_2\,;
\nonumber\\
\calC_2\equiv&
 - \frac{\kappa^2\mu(1-\lambda)}{1-3\lambda}\Delta\Pi^\calR
+ a^2\phi_0' \left( \delta{Z}_{\delta\phi}
 - \frac{2\kappa^2\mu}{1-3\lambda}\Delta\delta\phi \right)
 - 4\calH(3\alpha_1+8\alpha_2)\Delta^2\calR\,,
\label{calC2}
\end{align}
where we have introduced $\calC_2$ to denote the induced (secondary)
constraint.
The Poisson algebra of $\calC_2$ with respect to other
constraints $\calC_A$ and $\calC_B$ and with respect to $\calH^{(s)}$
 are easily found to be
\begin{align}
\left\{ \calC_A, \calC_2 \right\} \neq & 0 \, ,
\\
\left\{ \calC_B, \calC_2 \right\} = & 0 \, ,
\\
\left\{ \calH^{(s)}, \calC_2 \right\} \neq & 0 \, .
\end{align}
The exact expressions for these Poisson brackets are not necessary
as we shall see shortly.

At this point, adding the new constraint $\calC_2$, let us
consider a new constrained Hamiltonian,
\begin{equation}
\calH^{(s)}_\mathrm{total}=\calH^{(s)} + A\calC_A + B\calC_B + \lambda\calC_2 \, ,
\end{equation}
where $\lambda$ is the Lagrange multiplier associated with $\calC_2$.
Since $\calC_B$ commutes with the other constraints as well as with the
Hamiltonian $\calH^{(s)}$, it is a first class constraint.
As for $\calC_A$ and $\calC_2$, their consistency
with the equations of motion require
\begin{align}
\frac{d\calC_A}{dt} = & \frac{\partial\calC_A}{\partial{t}}
 + \int d^3y\Bigl[\left\{ \calC_A, \calH^{(s)}(y) \right\}
 + \lambda(y) \left\{ \calC_A, \calC_2(y) \right\}\Bigr] = 0 \, ,
\\
\frac{d\calC_2}{dt} = & \frac{\partial\calC_2}{\partial{t}}
 + \int d^3y\Bigl[\left\{ \calC_2, \calH^{(s)}(y) \right\}
 + A(y) \left\{ \calC_2, \calC_A(y) \right\}\Bigr] = 0 \, .
\end{align}
Since $\left\{ \calC_2, \calC_A \right\} \neq 0$, these two equations
determine $A$ and $\lambda$. Thus the two constraints $\calC_A=\calC_2=0$ are
second class. Hence in particular, $A$ is {\em not} a
gauge degree of freedom as Einstein gravity: $A$ is determined by
the consistency of the constraints with the equations of motion,
and thus the time reparametrization symmetry is lost.

Finally, let us discuss the gauge transformation properties of physical
quantities. Since there is no temporal gauge degree of freedom,
the only remaining gauge degree of freedom is the one associated with
spatial gauge transformations (of scalar type), and $\calC_B$ is
the generator of the spatial gauge transformations.
It is then not difficult to calculate the gauge transformations of
physical quantities. Since $\calC_B = \Pi^E$, a physical
quantity $X=X(\Pi^q,q)$, where $q=\{\calR,\delta\phi,E\}$,
will transform under a spatial gauge transformation induced
by $x^i\to\bar x^i=x^i-\partial^i\xi$ as
\begin{align}
\bar X=& X+\delta_gX \, ;
\nonumber\\
&\delta_gX = \left\{ X, \int d^3x \,\xi\,\calC_B \right\}
 = \frac{\partial X}{\partial E}\frac{\delta}{\delta\Pi^E}
 \int d^3x \,\xi\,\calC_B
= \frac{\partial X}{\partial E}\,\xi\, .
\end{align}
In particular, $\delta_gE=\xi$ and all the other canonical variables
are automatically gauge invariant.

\subsection{Hamiltonian reduction}
\label{subsec_no_proj_reduction}

Now we can see how many dynamical degrees of freedom are left after
 reducing the phase space. Since there are one first class and two
second class constraints, we have only one dynamical degree of freedom,
or one pair of canonical variable-conjugate momentum.\footnote{According to Ref.~\cite{Blas:2009yd}, at linear level
there exists a single extra degree of freedom which is manifest only around
spatially inhomogeneous and time-dependent background. It was argued that
the absence of extra degree of freedom in the non-projectable case
 is the artifact of considering linear perturbations around a homogeneous
background. In our opinion, the presence of a fixed background time-slicing seems
necessary to make the theory consistent, but it needs a
more careful analysis in order to clarify this issue,
which is beyond the scope of the present paper. }
 Following the
method developed by Faddeev and Jackiw~\cite{Faddeev:1988qp}, we
derive the reduced Hamiltonian by inserting the constraints
$\calC_A=\calC_B=\calC_2=0$ into the action in the first order form,
%
%
$\delta_2S^{(s)}=\int d^4x\calL_2^{(s)}$,
 with 
 $\calL_2^{(2)}$ given by (\ref{1st_order_L}).

We proceed as follows. We first use $\calC_B = 0$ to remove $\Pi^E$.
Since $\calH^{(s)}$ does not involve $E$, this automatically remove $E$
as well.
Next, by applying $\calC_A = 0$, we eliminate $\Pi^\calR$.
At this stage, $\calL_2^{(s)} = \calL_2^{(s)}(\calR,\delta\phi,\Pi^{\delta\phi})$
 and $\calC_2 = \calC_2(\calR,\delta\phi,\Pi^{\delta\phi})$.
Now, in place of $\delta\phi$ and $\Pi^{\delta\phi}$,
if we introduce auxiliary variables,
\begin{align}
\label{Q}
Q = & \delta\phi- \frac{\phi_0'}{\calH}\calR \, ,
\\
\label{Pi_Q}
Y = &
\Pi^{\delta\phi} - \frac{a^3}{\calH}\left( \frac{\phi_0'}{a} \right)'\calR \,,
\end{align}
we find that the action at this stage takes the form,
\begin{align}
\delta_2S_\mathrm{temp}^{(s)}=&\int d^4x
\left[YQ'- \calH_\mathrm{temp}^{(s)}(Q,Y,\calR)
-\lambda C_2(Q,Y,\calR)\right]\,;
\nonumber\\
&\calH_\mathrm{temp}^{(s)}(Q,Y,\calR) =
\calA(Y,Q) - \calB(Y,Q)\Delta\calR + \calR\calF\Delta\calR \, ,
\label{noproj_Htemp}
\end{align}
where $\calA$ and $\calB$ are, respectively,
quadratic and linear in $Y$ and $Q$, in the form,
\begin{eqnarray}
\calA(Y,Q) &= & \calA_1{Y}^2 + \calA_2YQ + Q\calA_3Q \, ,
\\
\calB(Y,Q) &= & \calB_1Y + \calB_2Q \, ,
\label{calABform}
\end{eqnarray}
 and $\calF$ is an operator quadratic in $\Delta$.
Finally, we use $\calC_2=0$ to express $\calR$ in terms of $Q$ and $Y$ as
\begin{equation}\label{noproj_solR}
\calR = \frac{1}{2}{\calF}^{-1}\calB \,.
\end{equation}
Then we see that $Y$ is indeed the canonical conjugate to $Q$, $Y=\Pi^Q$,
and we end up with
\begin{align}
\label{Qaction}
\delta_2S_\star^{(s)}
= & \int d^4x \left[ \Pi^QQ' - \calH_\star^{(s)}(\Pi^Q,Q) \right] \,,
\\
\label{noproj_H}
\calH_\star^{(s)} = & \calA(\Pi^Q,Q)
 - \frac{1}{4}\calB(\Pi^Q,Q)\frac{\Delta}{\calF}\calB(\Pi^Q,Q) \, .
\end{align}
The explicit forms of $\calA$, $\calB$ and $\calF$
in the above are given in Appendix \ref{appendix_A}.

It is interesting to note that the variable $Q$ introduced in (\ref{Q})
appears to be equal to the gauge invariant scalar field
perturbation on flat slicing in Einstein gravity for which $\calC_A$ is
a first class constraint, although in the present case
there is no physical meaning associated with the variable $Q$
since $\calC_A$ is not first class.

Using the Hamiltonian equation of $Q$,
we can eliminate $\Pi^Q$ in favour of $Q$ to write (\ref{Qaction})
purely in terms of $Q$ as
\begin{equation}\label{Qaction2}
\delta_{2}S_\star^{(s)}
 = \int d^4x \left\{Q'\frac{1}{4\calG_1}Q'
 + Q\left[ \left( \frac{\calG_2}{4\calG_1} \right)'
 + \frac{\calG_2^2}{4\calG_1} - \calG_3 \right]Q \right\} \, ,
\end{equation}
where
\begin{equation}
\calG_1 \equiv \calA_1 - \frac{\calB_1^2}{4\calF\Delta} \, ,
\quad
\calG_2 \equiv \calA_2 - \frac{\calB_1\calB_2}{2\calF\Delta} \, ,
\quad
\calG_3 \equiv \calA_3 - \frac{\calB_2^2}{4\calF\Delta} \, .
\end{equation}
If we change the variable by introducing
\begin{equation}
Q \equiv \sqrt{2\calG_1}u \, ,
\end{equation}
we can write (\ref{Qaction2}) as
\begin{equation}\label{uaction}
\delta_2S_\star^{(s)}
 = \int d^4x \frac{1}{2}\left\{ {u'}^2
 - u \left[ \left( \frac{\calG_1'}{2\calG_1} \right)'
 - \left(\frac{\calG_1'}{2\calG_1}\right)^2
 - \calG_1 \left( \frac{\calG_2}{\calG_1} \right)'
 - \calG_2^2 + 4\calG_1\calG_3 \right]u \right\} \, .
\end{equation}
By varying (\ref{uaction}) with respect to $u$, we obtain
the equation of motion of $u$ as
\begin{equation}\label{uequation}
u'' + \left[ \left( \frac{\calG_1'}{2\calG_1} \right)'
 - \left(\frac{\calG_1'}{2\calG_1}\right)^2
- \calG_1 \left( \frac{\calG_2}{\calG_1} \right)'
 - \calG_2^2 + 4\calG_1\calG_3 \right]u = 0 \, .
\end{equation}
In the limiting cases $k \to \infty$ and $k \to 0$, we
 obtain useful expressions: they are presented in Appendix \ref{appendix_B}.

\section{Scalar perturbation: With projectable condition}
\label{sec_proj}

\subsection{First order form of the Lagrangian}

If we first apply the projectable condition $A = A(\eta)$
to (\ref{reduced_2nd_scalar_action}), we can eliminate two
terms in (\ref{reduced_2nd_scalar_action}), which are of the form,
\begin{equation}
A \times (\text{spatial derivatives of other canonical variables}) \, .
\end{equation}
Since $A$ is now a function of time and thus these terms can be made
as total spatial derivatives,
and we can drop them from the beginning. Thus now we have
\begin{align}\label{2nd_action2}
\delta_2S^{(s)} = & \int d^4x a^2
\left\{ \frac{1-3\lambda}{\kappa^2} \left[ 3{\calR'}^2 -6\calH A\calR'
 + \left( \calH' + 2\calH^2 \right)A^2
- 2\calR'\Delta(B-E') \right]
\right.
\nonumber\\
& + \frac{1-\lambda}{\kappa^2} \left[ \Delta(B-E') \right]^2
 - 2\mu\calR\Delta\calR +\frac{2}{a^2}(3\alpha_1+8\alpha_2)(\Delta\calR)^2
\nonumber\\
& \left. + \frac{1}{2}{\delta\phi'}^2 - \delta{Z} -
\frac{1}{2}a^2V_{\phi\phi}\delta\phi^2 - \phi_0'A\delta\phi'
- 3\phi_0'\calR'\delta\phi -a^2V_\phi A\delta\phi
 + \phi_0'\delta\phi\Delta(B-E') \right\} \, .
\end{align}
Finding the conjugate momenta, we can see that this time we have
 different $\Pi^E$ as
\begin{equation}
\label{Pi_E2}
\Pi^E = a^2 \Delta \left[ 2\frac{1-3\lambda}{\kappa^2}\calR' -
2\frac{1-\lambda}{\kappa^2}\Delta(B-E') - \phi_0'\delta\phi \right] \, .
\end{equation}
Comparing with (\ref{Pi_E1}), we can see that there is no $\calH A$ term,
 which is already dropped out. $\Pi^\calR$ and $\Pi^{\delta\phi}$ are
 the same as (\ref{Pi_psi1}) and (\ref{Pi_deltaphi1}), respectively. Then,
 using (\ref{Pi_psi1}), (\ref{Pi_deltaphi1})
and (\ref{Pi_E2}) to replace time derivatives with the conjugate momenta,
after some arrangement, we find
\begin{align}
\label{proj_L}
\calL_2^{(s)} = & \Pi^\calR\calR' + \Pi^{\delta\phi}\delta\phi'
+ \Pi^EE' - \calH^{(s)} -\frac{3a^2(1-3\lambda)^2}{2\kappa^2}\calH^2A^2
 - A\calC_A - B\calC_B \, ,
\\
\calH^{(s)} = & \frac{\kappa^2}{4a^2} \left[- \Pi^\calR\Delta^{-1}\Pi^E
 + \frac{3}{2}\left(\Delta^{-1}\Pi^E\right)^2 +
\frac{2}{\kappa^2}{\Pi^{\delta\phi}}^2
+ \frac{1-\lambda}{2(1-3\lambda)}{\Pi^\calR}^2\right]
 + \frac{\kappa^2}{2(1-3\lambda)}\phi_0'\Pi^\calR\delta\phi
\nonumber\\
& + \frac{3\kappa^2a^2}{4(1-3\lambda)}{\phi_0'}^2\delta\phi^2
+ 2a^2\mu\calR\Delta\calR -2(3\alpha_1+8\alpha_2)(\Delta\calR)^2
 + \frac{a^2}{2} \left( 2\delta{Z} +a^2V_{\phi\phi}\delta\phi^2 \right) \, ,
\\
\calC_A = & \frac{3(1-\lambda)}{2}\calH\Pi^\calR
 + \phi_0'\Pi^{\delta\phi} -
\frac{3(1-3\lambda)}{2}\calH\Pi^E + a^2 \left( 3\calH\phi_0' + a^2V_\phi
\right)\delta\phi \, ,
\\
\calC_B = & \Pi^E \, ,
\end{align}
where again we have used (\ref{useful}) to simplify $A^2$ terms.

\subsection{Equations of motion}

From the Lagrangian (\ref{proj_L}), we can solve $A = A(\eta)$ to obtain
\begin{equation}\label{sol_A}
A(\eta)
= -\frac{\kappa^2}{3a^2(1-3\lambda)^2\calH^2}\,
\frac{\int d^3x\, \calC_A}{\int d^3x}\, .
\end{equation}
Here we should note that $\calC_A$ is linear in the perturbation variables.
Hence its integral over space\footnote{Note that in this version of HL gravity, the Hamiltonian constraint, given by the integration over the whole
space, may give rise to an extra dark-matter-like component in the
Friedmann equation (\ref{friedmann_eq}), if a non-trivial spatial
 boundary is considered, as pointed out in Ref.~\cite{Mukohyama:2009mz}.}
singles out zero modes or spatially homogeneous
modes. Thus, if we consider an infinite spatial volume,
the integral vanishes because spatially homogeneous modes are not
included in perturbation by construction\footnote{We may include
spatially homogeneous modes of the canonical variables. However,
they simply describe a global gauge degree of freedom corresponding
to the global time reparametrization given by (\ref{sol_A}).
}.
 Therefore
\begin{equation}
A(\eta)=0\,.
\end{equation}
This again means that there is no time reparametrization symmetry,
but the situation is different from the case without the projectable
condition. Previously $A$ was dependent on and determined by the constraints,
but now $A$ simply vanishes: there is only a single way of time-slicing.
This is analogous to the concept of ``absolute time'' in Newton gravity.
Meanwhile, $B$ is a Lagrange multiplier and its equation of motion,
$\calC_B = \Pi^E = 0$, is the momentum constraint, representing the spatial
gauge degree of freedom again. Then once again all the other canonical
variables other than $E$ are gauge invariant.

Now following Ref.~\cite{Faddeev:1988qp},
we may plug $A=\Pi_E=0$ into the action to obtain
the reduced action for true dynamical variables.
 Denoting by a subscript $\star$ the Lagrangian
with $A=\Pi^E=0$ substituted, we find
\begin{align}
\calL_{2\star}^{(s)} = & \Pi^\calR\calR' + \Pi^{\delta\phi}\delta\phi'
 - \calH_\star^{(s)} \,,
\label{reducedL2}\\
\calH_\star^{(s)} = & \frac{{\Pi^{\delta\phi}}^2}{2a^2}
 +\frac{\kappa^2(1-\lambda)}{8a^2(1-3\lambda)}{\Pi^\calR}^2
 +\frac{\kappa^2}{2(1-3\lambda)}\phi_0'\Pi^\calR\delta\phi +
\frac{3\kappa^2a^2}{4(1-3\lambda)}{\phi_0'}^2\delta\phi^2
\nonumber\\
& + 2a^2\mu\calR\Delta\calR
 - 2(3\alpha_1+8\alpha_2)(\Delta\calR)^2
+ \frac{a^2}{2}\left( 2\delta{Z} + a^2V_{\phi\phi}\delta\phi^2 \right) \, .
\label{reducedH2}
\end{align}
Before proceeding further, we pause for the moment and
consider the number of remaining degrees of freedom.
As can be easily seen, there remains no further
constraint in (\ref{reducedH2}). Thus there are four degrees
of freedom in terms of canonical variables, or two configuration space
variables~\cite{2scalardof}. Namely, it is impossible
to reduce the perturbation degrees of freedom to a single degree
of freedom like $Q$ as before.

Given the reduced Hamiltonian density $\calH_\star^{(s)}$,
we can write down the Hamilton equations of motion for
the canonical variables to obtain
\begin{align}
\calR' = & \frac{\kappa^2(1-\lambda)}{4a^2(1-3\lambda)}\Pi^\calR
+ \frac{\kappa^2}{2(1-3\lambda)}\phi_0'\delta\phi \, ,
\label{psi_eq}
\\
{\Pi^\calR}' = & -2 \left[ 2a^2\mu\Delta\calR
- 2(3\alpha_1+8\alpha_2)\Delta^2\calR \right] \, ,
\label{Pipsi_eq}
\\
\delta\phi' = & \frac{\Pi^{\delta\phi}}{a^2} \, ,
\label{deltaphi_eq}
\\
{\Pi^{\delta\phi}}'
= &- \frac{\kappa^2}{2(1-3\lambda)}\phi_0'\Pi^\calR
 - \frac{3\kappa^2a^2}{2(1-3\lambda)}{\phi_0'}^2\delta\phi
 - a^2 \left( 2\delta{Z}_{\delta\phi} + a^2V_{\phi\phi} \right) \delta\phi \, .
\label{Pideltaphi_eq}
\end{align}
Combining these equations, we may eliminate $\Pi^\calR$ and
$\Pi^{\delta\phi}$ to obtain coupled second order differential equations
for $\calR$ and $\delta\phi$,
\begin{align}
\calR'' + 2\calH\calR' + \frac{\kappa^2(1-\lambda)}{a^2(1-3\lambda)}
 \left[ a^2\mu\Delta\calR - (3\alpha_1+8\alpha_2)\Delta^2\calR \right]
 = & \frac{\kappa^2}{2a^2(1-3\lambda)} \left( a^2\phi_0'\delta\phi \right)' \, ,
\\
\delta\phi'' + 2\calH\delta\phi'
 + \frac{\kappa^2}{2(1-3\lambda)}{\phi_0'}^2\delta\phi
 + \left( 2\delta{Z}_{\delta\phi} + a^2V_{\phi\phi} \right)\delta\phi
 = &- \frac{2\phi_0'}{1-\lambda}\calR' \, .
\end{align}
Note that the above equations may be obtained by eliminating
the canonical momenta from the Lagrangian (\ref{reducedH2}) by
using (\ref{psi_eq}) and (\ref{deltaphi_eq}).
If we do this, we find the kinetic part of the Lagrangian becomes
\begin{equation}
\calL_{2\star}^{(s)}
= \frac{2a^2(1-3\lambda)}{\kappa^2(1-\lambda)}{\calR'}^2
 + \frac{a^2}{2}{\delta\phi'}^2 + \cdots \, .
\end{equation}
This suggests that there is a ghost in the theory for $\lambda$
in the range $1/3 < \lambda < 1$.

Before concluding this section, it may be worth discussing
the special cases of $\lambda=1/3$ and $\lambda=1$.
In the case of $\lambda=1/3$, we see from (\ref{reducedH2}) that
we have an additional constraint,
\begin{eqnarray}
\Pi^\calR+3\phi_0'\delta\phi=0 \, .
\end{eqnarray}
Then if we eliminate $\Pi^\calR$ from the action, $\calR$
ceases to be dynamical, and $\delta\phi$
becomes the only remaining dynamical degree of freedom.

In the case of $\lambda=1$, the ${\Pi^\calR}^2$ term in the Lagrangian
vanishes. This means if we go back to the second order form of the
Lagrangian, we cannot eliminate $\Pi^\calR$. Namely, we have
\begin{eqnarray}
\left.\calL_{2\star}^{(s)}\right|_{\lambda=1}
&=&a^2\frac{\delta\phi'{}^2}{2}
+\frac{3\kappa^2a^2}{8}{\phi_0'}^2\delta\phi^2
-\frac{a^2}{2}\left( 2\delta{Z} + a^2V_{\phi\phi}\delta\phi^2 \right)
\nonumber\\
&& - 2a^2\mu\calR\Delta\calR
 + 2(3\alpha_1+8\alpha_2)(\Delta\calR)^2
 +\Pi^\calR\left(\calR'+\frac{\kappa^2}{4}\phi_0'\delta\phi\right)\,,
\end{eqnarray}
and $\Pi^\calR$ remains as a Lagrange multiplier for the
constraint $\calR'+\kappa^2\phi_0'\delta\phi/4=0$.
Eliminating $\delta\phi$ by using the constraint gives a Lagrangian
for $\calR$ which contains ${\calR''}^2$. Thus the system
gives a fourth-order differential equation for $\calR$.
In other words, there is no change
in the number of dynamical degrees of freedom in this case.
Whether one of them is a ghost is an issue that needs a more
detailed analysis, which is out of the scope of the present paper.

\section{Conclusion}
\label{sec_conclusion}

In this paper, we formulated the linear cosmological perturbations
in HL gravity. The complication is that the space-time structure of
HL gravity is very different from that of Einstein gravity because of
the lack of general covariance and the consequent projectability.
 This in turn means that the issue of gauge transformations for the
cosmological perturbations is also different. Therefore, we first of all
 have to address this question to properly extract true dynamical degrees
of freedom and to study their evolution with the equations of motion.
We studied this subject in the case when the matter sector is given
by a canonical scalar field.

To systematically reduce the number of degrees of freedom,
we employed the Hamiltonian formalism and derived the Lagrangian
in the first order form, the analysis of which immediately tells us
the true dynamical degrees of freedom.
We found that irrespective of projectability, the tensor
perturbations have two independent degrees of freedom, or two polarizations,
 while the vector perturbations are not dynamical.
This is the same as in the case of Einstein gravity.
 For the scalar perturbations, however, the result depends on whether we
 apply the projectable condition or not.

When the projectability is not applied,
the Lagrangian looks similar to that of Einstein gravity where there
are two constraints corresponding to Hamiltonian and (scalar-type)
momentum constraints. But unlike Einstein gravity, the consistency of
the two constraints gives rise to a secondary constraint, and this
 new constraint and the Hamiltonian constraint become second class,
while the momentum constraint remains first class. Thus we lose
 the time reparametrization symmetry, and we are left with
a single dynamical variable (in configuration space).
It may be noted that mathematically this new constraint works
exactly like a gauge fixing condition. In this sense,
HL gravity without projectability is like Einstein gravity
but with a preferred time slicing.

With the projectability condition, we have an absolute time in the
 sense that time-slicing is apriori completely fixed irrespective of
the dynamics. Then the scalar gravitational degree of freedom, which
would be constrained in Einstein gravity, becomes dynamical.
Thus we are left with two independent degrees of freedom,
one from gravity and one from the scalar field.
We obtained their coupled second order different equations.

\subsection*{Acknowledgement}

We thank Bin Hu, Shinji Mukohyama, Mu-In Park and David Wands for
useful comments and discussions.
We would like to thank an anonymous referee for several comments.
JG is grateful to the Kavli Institute
 for Theoretical Physics China for hospitality during the program
 ``Connecting Fundamental Physics with Observations'', where this work
 was initiated and KIAS for hospitality where some part of this work was finished.
 JG is partly supported by a VIDI and a VICI Innovative
 Research Incentive Grant from the Netherlands Organisation for Scientific
Research (NWO).
 SK is supported by the National Research Foundation of Korea Grant funded by the
Korean Government [NRF-2009-353-C00007].
MS is supported in part
by JSPS Grant-in-Aid for Scientific Research (A) No.~21244033,
by JSPS Grant-in-Aid for Creative Scientific Research No.~19GS0219,
and by MEXT Grant-in-Aid for the global COE program at Kyoto University,
``The Next Generation of Physics, Spun from Universality and Emergence''.


\appendix

\section{Various functions introduced in Sec.~\ref{subsec_no_proj_reduction}}
\label{appendix_A}

In this appendix, we present the functions $\calA(\Pi^Q,Q)$, $\calB(\Pi^Q,Q)$
 and $\calF$ which appear in Section~\ref{subsec_no_proj_reduction}:
\begin{align}
\calA = & \calA_1{\Pi^Q}^2 + \calA_2\Pi^QQ + Q\calA_3Q \, ,
\\
\calB = & \calB_1\Pi^Q + \calB_2Q \, ,
\\
\calF= & \calF_1 + \calF_2\Delta + \calF_3\Delta^2 \, ,
\end{align}
where
\begin{align}
\calA_1 = & \frac{\kappa^2(1-\lambda)}{8a^2(1-3\lambda)}
 \left( \frac{\phi_0'}{\calH} \right)^2 + \frac{1}{2a^2} \, ,
\\
\calA_2 = & -\frac{\kappa^2(1-\lambda)}{4a^2(1-3\lambda)}
 \left( \frac{\phi_0'}{\calH} \right) \frac{a^3}{\calH}
\left( \frac{\phi_0'}{a} \right)' - \frac{\kappa^2}{2(1-3\lambda)}\calH
 \left( \frac{\phi_0'}{\calH} \right)^2 \, ,
\\
\calA_3 = & \frac{\kappa^2(1-\lambda)}{8a^2(1-3\lambda)}
 \left[ \frac{a^3}{\calH} \left( \frac{\phi_0'}{a} \right)' \right]^2
 + \frac{\kappa^2}{2(1-3\lambda)}\calH \left( \frac{\phi_0'}{\calH} \right)
 \frac{a^3}{\calH} \left( \frac{\phi_0'}{a} \right)'
\nonumber\\
& + \frac{a^2}{2} \left[ \frac{3\kappa^2}{2(1-3\lambda)}{\phi_0'}^2
 + a^2V_{\phi\phi} \right] - a^2 \left( \xi_3\Delta^3 - \xi_2\Delta^2
+ \xi_1\Delta \right) \, ,
\\
\calB_1 = & -\frac{\kappa^2\mu(1-\lambda)}{(1-3\lambda)\calH}
\frac{\phi_0'}{\calH} \, ,
\\
\calB_2 = & \frac{(1-\lambda)\kappa^2\mu}{(1-3\lambda)\calH}
\frac{a^3}{\calH} \left( \frac{\phi_0'}{a} \right)'
+ \frac{2\kappa^2\mu a^2}{1-3\lambda}
\left( \frac{\phi_0'}{\calH} \right)
+ 2a^2 \left( \frac{\phi_0'}{\calH} \right)
\left( \xi_3\Delta^2 - \xi_2\Delta + \xi_1 \right) \, ,
\\
\calF_1 = & -a^2 \left( \frac{\phi_0'}{\calH} \right)^2
\left( \xi_1 + \frac{\kappa^2\mu}{1-3\lambda} \right) \, ,
\\
\calF_2 = & a^2 \left( \frac{\phi_0'}{\calH} \right)^2 \xi_2
+ \frac{2(1-\lambda)\kappa^2\mu^2a^2}{(1-3\lambda)\calH^2}
- 2(3\alpha_1+8\alpha_2) \, ,
\\
\calF_3 = & -a^2 \left( \frac{\phi_0'}{\calH} \right)^2 \xi_3 \, .
\end{align}

\section{UV and IR limits of (\ref{uequation})}
\label{appendix_B}

Here we consider the ultraviolet and infrared limits of (\ref{uequation}).
If we schematically write (\ref{uequation}) as
\begin{equation}
u'' + \omega^2u = 0 \, ,
\end{equation}
we can easily find that in the ultraviolet limit,
\begin{equation}
\omega^2 \underset{k\to\infty}{\longrightarrow}
\left\{ \frac{4(1-\lambda)\kappa^2\mu^2}{(1-3\lambda){\phi_0'}^2}
 - \left[ \frac{(1-\lambda)\kappa^2}{a^2(1-3\lambda)}
 + \frac{4}{a^2}\left( \frac{\calH}{\phi_0'} \right)^2 \right]
 \left( 3\alpha_1+8\alpha_2 \right) \right\} \Delta^2 \, .
\end{equation}
Note that $\Delta^3$ terms precisely cancel each other~\cite{noproj_pert}.
 Meanwhile, in the infrared limit we have
\begin{equation}
\omega^2 \underset{k\to0}{\longrightarrow}
 \frac{2\kappa^2\mu}{1-3\lambda}\Delta - \frac{z''}{z} \, ,
\end{equation}
with $z \equiv a\phi_0'/\calH$.
If in addition we set $\kappa^2\mu = 1$ and $\lambda = 1$,
(\ref{uequation}) reduces to the well-known
perturbation equation in Einstein gravity~\cite{linear,reduction}
\begin{equation}
u'' - \Delta u - \frac{z''}{z}u = 0 \, .
\end{equation}

\end{document}